\documentclass[10pt]{iopart}
\pdfoutput=1
\usepackage[pdftex]{graphicx,color} 
\usepackage[T1]{fontenc}
\usepackage{caption}
\usepackage{float}
\begin{document} 

\title{Limits of light-trapping efficiency of prototypical
  lamellar 1-d metal gratings for amorphous silicon PV cells}

\author{D. I. Gablinger$^{1,*}$ and R.H. Morf$^1$}
\address{$^1$
Condensed Matter Theory Group\\
Paul Scherrer Institute\\
CH-5232 Villigen}
%\email{$^{*}$david.gablinger@psi.ch}

\begin{abstract}
One-dimensional lamellar gratings allow a particularly efficient way
for solving Maxwell's equations by expanding the electromagnetic
field in the basis of exact eigenmodes of the Helmholtz equation.
Then, the solution can be expressed analytically as a
superposition of these eigenmodes and the accuracy depends only on the
number of modes $N$ included. On this basis, we compute ideal limits
of light-trapping performance for prototypical lamellar metal surface relief gratings
in amorphous silicon (a-Si) PV cells assuming that light absorption in
the metal and front surface reflection can be suppressed. 
We show that geometric asymmetry can increase
absorption. For large enough $N$, convergence of absorption spectra
for E polarisation is reached. For H polarisation it is reached for
wavelengths $\lambda<$680-700 nm,  while the integrated AM1.5-weighted
absorption varies by less than 1\% at large $N$. For an  a-Si layer with
height 200 nm and normal incidence, we obtain upper limits of the total absorption of
79\% for E-polarisation with asymmetric- and  90\% for H-polarisation
with sine-like lamellar grating reflectors, whereas a planar reflector
yields  a limit of 62\%,  where 100\% stands for complete absorption
for $350<\lambda<$770 nm.
\end{abstract}
\ioptwocol

%%%%%%%%%%%%%%%%%%%%%%% References %%%%%%%%%%%%%%%%%%%%%%%%%
%\bibliographystyle{osajnl}
%\bibliography{referenzen}
%%%%%%%%%%%%%%%%%%%%%%%%%%  body  %%%%%%%%%%%%%%%%%%%%%%%%%%

\section{Introduction}

In this paper, we examine fundamental upper limits of the light
trapping efficiency of a class of one-dimensional gratings consisting
of amorphous silicon (a-Si) and silver as a reflecting substrate. We focus
uniquely on the {\it light-trapping properties} of a class of lamellar
surface relief
gratings for which we can solve Maxwell's equations in the continuum
limit, i.e. without discretization in physical space, and use
analytical expressions for the electromagnetic field as a function of
the spatial coordinates. The main numerical work consists in the
solution of algebraic eigenvalue problems, and at the end of solving a
system of linear equations. Everything else can be expressed analytically.

In order to obtain upper limits for light-trapping efficiency we use an
ideal absorption-free reflector made from a hypothetical metal,
idealised silver (Ag$^*$) whose dielectric permittivity is real and
identical to the real part of the permittivity of silver
$\epsilon_{Ag}(\lambda)$. We also employ an ideal
antireflection coating (AR$^*$) that suppresses reflection in the
whole wavelength range of a-Si absorption. The reason for this choice
of back reflector and antireflection coating, is that we wish to focus
our attention uniquely on the light-trapping efficiency and want to
make sure that all the photons in the reflected light have had a
chance of being absorbed in the semiconductor. 

We wish to point out that the absorption in a real metal will not only lead 
to a reduction of absorption
in the photoelectrically active region of the semiconductor, but also
affect the number of photons leaving the cell and in this way reduce  the
intensity of reflected light. Our results present the basis for predicting the
improvement possible by reducing the parasitic absorption in the
metallic reflector.
In the future, we may hope that losses by absorption in the metal
reflector and reflection at the front surface will be reduced more
than presently possible or even will be overcome by some novel methods.

Such exact limits have traditionally played an important role,
the best known example being the Carnot limit of thermodynamic
engines, as well as in photovoltaic converters, e.g. the detailed
balance limit established by Shockley and Queisser
\cite{Shockley:1961lr}, or the famous
limit obtained for random scatterers by Yablonovitch and Cody
\cite{Yablonovitch:1982lr}. These upper limits serve as a yardstick against which
progress can be measured and which indicate what might become
possible.% if we are inventive.

In the context of light-trapping induced by diffraction gratings, we are
not able to present upper limits of such general validity. They have
been the subject of recent work by
\cite{Yu12102010,haug2011resonances,Naqavi2011optimal1d}. They are
typically made within mode-coupling theory using assumptions about
possible couplings between the modes that are consistent with general
principles, e.g. detailed balance. 

Here, we wish to present limits of light-trapping efficiency for concrete
diffraction gratings without making assumptions about the couplings
between modes, but obtaining them from exact calculations which are
possible for the classes of diffraction gratings  with lamellar
geometry chosen for the present study as well as generalizations of
those with 2-dimensional periodicity.

All work on light-trapping is motivated by the need to increase light
absorption in solar cells with indirect
bandgap semiconductors. The use of diffraction gratings
to improve absorption in solar cells is well known, and has been
described in detail in the review article by S. Mokkapati and K.R.
Catchpole\cite{mokkapati2012nanophotonic}.

We present an optical study of diffraction structures with a-Si
as semiconductor material primarily for practical reasons.
In particular, silicon is abundant, and methods for vapour deposition
of a-Si are well developed. Unlike thin solar cells
using a direct band-gap semiconductor that absorbs very well for all
photon energies above the band-gap, a-Si becomes a weak
absorber in the near-infrared, and thus long optical
path lengths within the material are required. On the other hand,
electronic properties of the semiconductor such as recombination losses
set upper limits on layer thickness, and a compromise between absorption
and these electronic properties is usually found at around $200$
nm layer thickness for a-Si solar cells. All results
for the spectrally integrated absorption $A_{int}$ quoted in this
paper, represent spectrally integrated values weighted with the solar AM1.5
spectral photon density. Note that the number of absorbed photons
is the relevant quantity, when the quantum
effciency is limited to one electron-hole pair per absorbed photon.

Our method for solving Maxwell's equations follows reference \cite{morf1995exponentially}.
The method described therein solves the diffraction problem for a
lamellar gratings by first solving the Helmholtz eigenvalue equation for
each layer. For each layer, there is a set of eigenvalues and
eigenfunctions that form the basis in which the electromagnetic field
is expanded. Continuity of the field and its derivative normal to the
layer, see below, allows to relate expansion coefficients between
different layers, and at the bottom and top of the structure there exist
boundary conditions that define a unique solution to the system of
linear equations linking all expansion parameters.
The eigenmodes obtained with this method are numerically exact, but
of course finite in number.

Following the strategy outlined in reference \cite{morf1995exponentially},
the boundary value problem can be solved for arbitrary grating depth
and arbitrary number of eigenmodes without encountering numerical
instability. However, the finite number of eigenmodes
taken into account implies that the resulting calculation is still approximate.
Interfaces with sharp optical boundaries with discontinuities in the
permittivities are inherent to the problem and lead to the excitation
of an infinite number of evanescent modes with very short decay length
in the perpendicular direction. Therefore high spatial resolution or,
as in our case, a large number of modes need to be taken into account.
The case of the polarisation with magnetic field parallel to the grating
grooves (H polarisation) is particularly slowly converging as the
H field has discontinuities in its normal derivative at interfaces
between different materials. The case of E polarisation has better
convergence properties as the first derivative of the E field is continuous,
only its second derivative being discontinuous. These discontinuities
of first (H-polarisation) or second (E-polarisation) derivatives are particularly large
at interfaces between metals and semiconductor, in our case between
silver and a-Si. This difficulty needs to be addressed
in particular in the spectral range where the absorptivity of the
semiconductor becomes weak, i.e. where light-trapping is most needed.
It is this problem that we address in the present work.
\begin{figure}[ht]
  \centering
  \includegraphics[width=6.6cm]{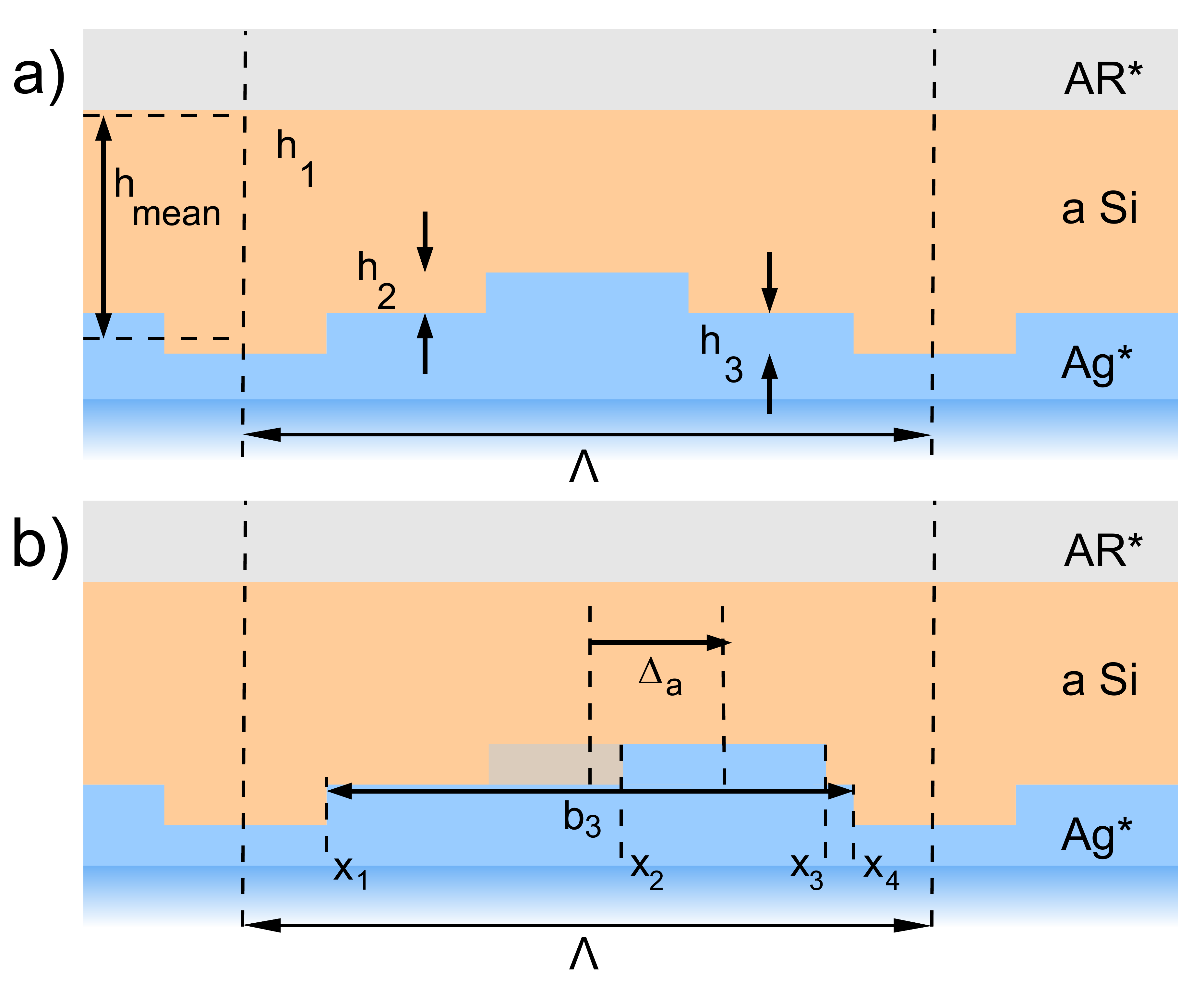}
  \vskip -2mm 
  \caption{Grating structure with two grating layers: a) Fully symmetric,
    b) Mirror axis in upper layer shifted.
    Mirror symmetry in individual layers remains.}
  \label{sym-antisym}
\end{figure}

We limit our studies to perpendicular incidence and %in this publication
we only treat the absorption in the semiconductor layer, and note
that the parasitic absorption in the metal layer will be discussed
in a forthcoming publication. 
Furthermore, we confine our discussion
to structures with two interfaces per layer, leading to a mirror symmetry
in each grating layer, as shown in Figure \ref{sym-antisym}. We benefit
from this symmetry property to separate the eigenfunction space into
symmetric and antisymmetric subspaces in each layer. As long as the
entire structure retains the same mirror plane, then symmetric eigenfunctions
do not couple to antisymmetric eigenfunctions. Since the zero order
light incident perpendicular to the device belongs to the symmetric
class, only symmetric eigenfunctions will couple to the incident light.
As a coupling is induced between symmetric and antisymmetric eigenfunctions
in different layers by asymmetric structures of the type shown in
the lower panel of Figure \ref{sym-antisym}, it is interesting to
study whether asymmetric structures lead to improved light trapping,
cf. Reference \cite{heine1995submicrometer}. 
The simplifications made possible by geometrical symmetry and
perpendicular incidence allow us to present results corresponding to
resolutions that are a factor of two larger. We present our
calculations with varying resolutions to provide a measure of the
precision of such calculations and to assess the continuum limit.

In section 2, we outline the problem and our approach for its solution.

In section 3.1 we examine the absorption behavior in the semiconductor
layer in a one-layer grating with rectangular geometry. Specifically,
we look at a structure with an absorption free metallic substrate
and an absorbing semiconductor layer with an antireflective cover.
In particular we describe how the absorption extends to larger wavelengths
as a function of grating depth, and we find an optimum depth for maximum
absorption. This optimum depth is almost the same for both polarisations.

In Section 3.2 we investigate multi-layer staircase gratings that
mimic sinusoidal gratings. Here we observe again that the optimum
thickness for the grating structures does not differ much between
the two polarisations. We note that with increasing number of grating
layers for the same depth and structure, the spectra differ in some
parts, while the integrated absorption $A_{int}$ is almost independent of the
number of grating layers included. The absorption depends little on
the period $\Lambda$ of the structure for both a rectangular and
a stair-case sine grating.

In Section 3.3, we investigate asymmetric structures. In the simplest
case, they consist of two symmetric layers with interfaces between
the metal and the semiconductor. As layers are moved horizontally
with respect to one another, symmetry of the entire structure is broken,
as shown in Figure \ref{sym-antisym} b). As long as the symmetry
axes for the two layers coincide, only the symmetric eigenfunctions
are populated in each layer. But as soon as the symmetry is broken,
symmetric eigenfunctions couple to antisymmetric eigenfunctions of different layers, and vice versa, such that all modes are populated. We investigate how these additional modes
excited by the coupling between symmetric and antisymmetric eigenfunctions
may increase absorption. Note that these symmetry considerations apply
only to the case of perpendicular incidence.%, which is the topic of
%this paper.

The paper closes with a discussion of the observed effects and an
outlook.

\section{Method}

In this section we first give a brief recapitulation of the calculational
method \cite{morf1995exponentially}. Then we discuss the specific
changes necessary for the present calculations.

\paragraph*{Setup}

\begin{figure}[ht]
  \centering
  \includegraphics[width=7cm]{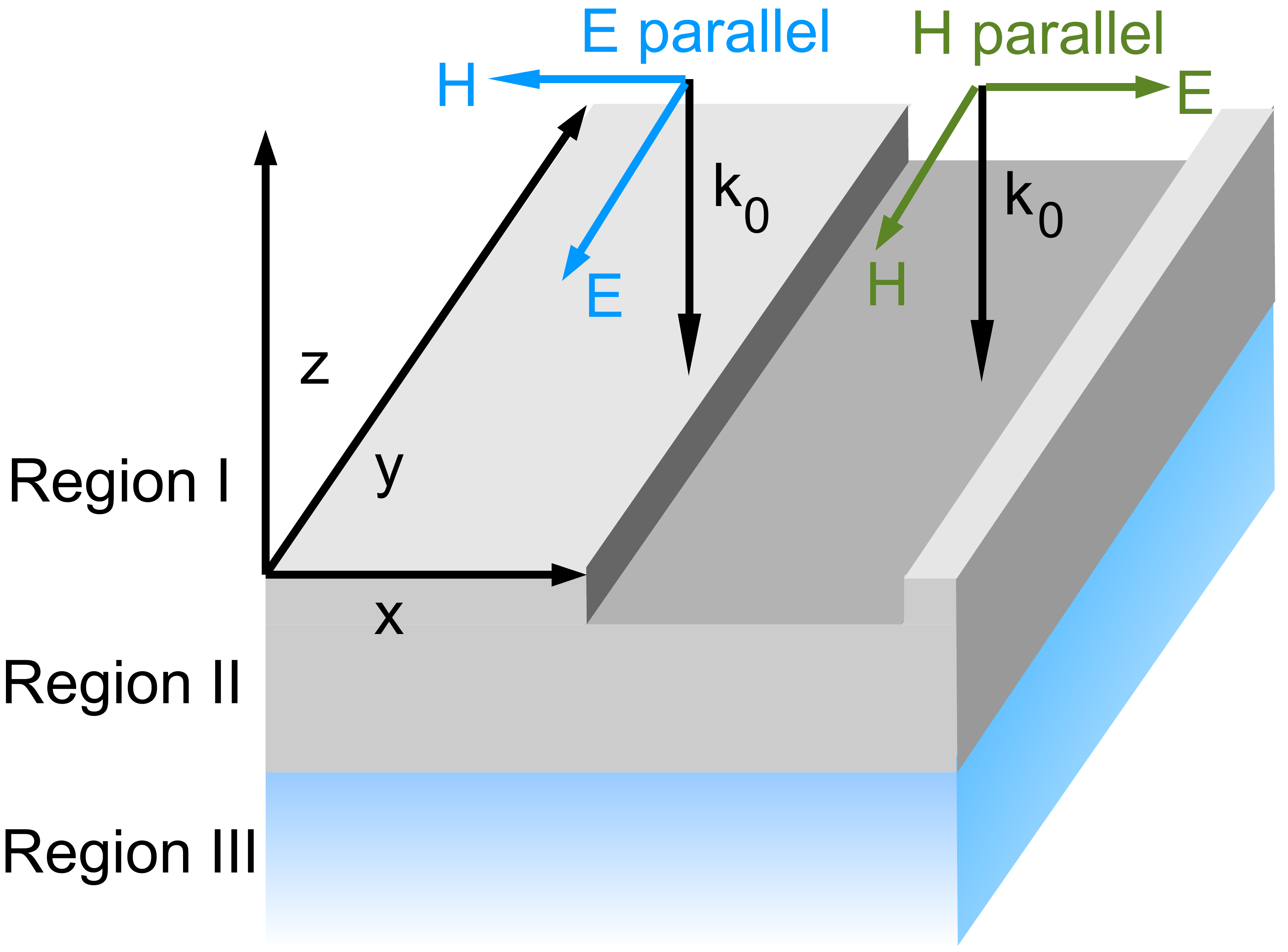}
  \vskip -3mm
  \caption{\label{coordinates} General problem setup and polarisation convention.}
\end{figure}

We split the diffraction problem into regions I-III, as shown in Figure
\ref{coordinates}, using the polarisation convention denoted there.
We denote $F(x,y)$ as either the electric field in the y-direction
$E_{y}(x,y)$ for E parallel polarisation or similarly the magnetic
field in the y-direction $H_{y}(x,y)$ for H parallel polarisation.
In region I, we study the case of perpendicular incidence, that is
the incident field can be written as 
\begin{equation}
F^{inc}(x,z)=1\times\exp(-ik_{0}z),\label{F1}
\end{equation}
where $k_0=2\pi/\lambda$ and for regions I and III, the outgoing light can be written as
a superposition of plane waves,  
\begin{eqnarray}
F^{I}(x,z) & = & \sum_{n=-M}^{M}a_{n}^{+(I)}\exp(i(\alpha_{n}x+\chi_nz)),\\
F^{III}(x,z) & = & \sum_{n=-M}^{M}a_{n}^{-(III)}\exp(i(\alpha_{n}x-\chi_nz)),
\end{eqnarray}
where 
\begin{equation}
\alpha_{n}=\frac{2\pi}{\Lambda}n,\qquad\textrm{and}\qquad\chi_{n}=\sqrt{\epsilon^{I,III}k_{0}^{\,2}-\alpha_{n}^{\;2}}.
\end{equation}
Region II contains the diffraction grating consisting of $q$ layers,
with thickness $h_{j}=z_{j+1}-z_{j}$, where $1\le j\le q$ refers
to the layer position within the grating.

\paragraph*{Grating Problem}
We briefly sketch our method for solving the grating problem,
developed by one of us \cite{morf1995exponentially}.
As we confine our study to perpendicular incidence, we note that the
fields $E$ and $H$ do not depend on y. Within Region II, we briefly
note that because of the lamellar structure, inside each layer the
permittivity $\epsilon(x,z)$ does not depend on the $z$. Thus, a
separation of variables becomes possible and the field $F(x,z)$ can
be written as a sum of products $X_{n}(x)\times Z_{n}(z)$, the functions
$X_{n}(x)$ is an eigenfunction Helmholtz equation which needs to
be solved within each grating layer. Because $\epsilon(x)$ is piecewise
constant, the Helmholtz equation can be written piecewise, too. For
simplicity, we limit ourselves to two domains of different permittivity:
\begin{eqnarray}
\partial_{x}^{2}\, X_{n}^{(j,1)}(x)+k_{0}^{\;2}\epsilon_{1}^{(j)}X_{n}^{(j,1)}(x) & = & \mu_{n}^{(j)2}X_{n}^{(j,1)}(x)\label{eq:helmholtz1}\\
\partial_{x}^{\;2}X_{n}^{(j,2)}(x)+k_{0}^{\;2}\epsilon_{2}^{(j)}X_{n}^{(j,2)}(x) & = & \mu_{n}^{(j)2}X_{n}^{(j,2)}(x)\label{eq:helmholtz2}
\end{eqnarray}
Here, equation \ref{eq:helmholtz1} refers to domain 1 and equation
\ref{eq:helmholtz2} refers to domain 2 and $\mu_{n}^{(j)^2}$ is $n$-th
eigenvalue of the Helmholtz equation in layer $j$ and
$X_{n}^{(j,1)}(x)$ and $X_{n}^{(j,2)}(x)$ are the domain-wise parts of the
the corresponding $n$-th eigenfunction $X_{n}^{(j)}(x)$
They need to fulfill a set of boundary conditions
at interfaces $x_{i}$ in addition to the eigenvalue equations, 
\begin{eqnarray}
X_{n}^{(j,1)}(x_{i}^{-}) & = & X_{n}^{(j,2)}(x_{i}^{+})\\
\frac{1}{\epsilon_{1}^{(j)}}\partial_{x}X_{n}^{(j,1)}(x)\mid_{x=x_{i}-\delta} & = & \frac{1}{\epsilon_{2}^{(j)}}\partial_{x}X_{n}^{(j,2)}(x)\mid_{x=x_{i}+\delta}
\end{eqnarray}
for $H$ parallel polarised light in the limit $\delta\rightarrow0$,
and likewise for $E$ parallel polarised light by replacing $\epsilon_{1}$
and $\epsilon_{2}$ by $1$. 

The electromagnetic field $F^{(j)}(x,z)$ can then be expanded in terms
of the eigenfunctions $X_n^{(j)}(x)$
as a sum of modes propagating in between the adjacent layers,
\begin{eqnarray}
F^{(j)}(x,z)& = &\sum_{n=1}^{N}\left(a_{n}^{(j)-}/A_n^{(j)}(z)+
  a_{n}^{(j)+}A_n^{(j)}(z)\right)X_{n}^{(j)}(x),\nonumber \\
A_n^{(j)}(z)& = &\exp(i\mu_{n}^{(j)}(z-z^{(j)}).\label{eq:planewave}
\end{eqnarray}
where $\mu_{n}^{(j)}$ is the square root
of the eigenvalue $\mu_{n}^{(j)2}$,
with the sign convention that the imaginary part of $\mu_n^{(j)}$ be
positive if it is non-zero. This sign definition ensures that the
amplitudes $a^{+}$ ($a^{-}$) refer to eigenmodes
that as functions of $z$ are exponentially decreasing (increasing),
respectively.

The Helmholtz equation together with the sets of basis functions and
the additional constraint of boundary conditions between the domains
define the numerical eigenvalue problem for each grating layer $j$.
This method allows the calculation of a freely selectable number of eigenvalues
and eigenfunctions to a precision limited only by machine number size:
The electromagnetic field is infinitely many times differentiable
inside each domain of constant permittivity, a fact that can be verified
easily by repeated differentiation of the Helmholtz equation. If the
functions $X^{(j,k)}(x)$ are expanded in terms of polynomials $P_{m}^{(j,k)}(x)$
that are orthogonal for $x$ inside the domain $D^{(j,k)}$, one obtains
exponential convergence in the number $M$ of included basis functions
for eigenvalues and eigenfunctions. Thus, within
a grating layer, our eigenmodes are obtained to the same precision
as by analytic calculation following references by Botten et
al. \cite{botten1981dielectric,botten1981finitely,botten1981highly}.
Note that the symmetry at perpendicular incidence together with the
symmetry within the grating layer results in either purely symmetric
or antisymmetric functions. If the basis functions $P_{m}^{(j,k)}(x)$
are also either symmetric( antisymmetric) for $m$ even(odd) or odd
$m$), half the number of coefficients defining the eigenfunctions
$X^{(j,k)}(x)$ will vanish, and therefore the symmetric eigenvalue
problem can be formulated either for symmetric or antisymmetric case
in terms of the $M/2$ non-zero variables and requires only
$\frac{1}{8}$ of the numerical effort as it
scales with the number of variables to the third power. For entirely
symmetric structures, this property carries over to the solution of
the radiation condition boundary value problem.

At the interface $z=z_{0}$ between grating layers, the field $F=E_{y}$
or $F=H_{y}$ has to satisfy the following continuity equations
\begin{eqnarray}
F(x,z_{0}+\delta) & \equiv & F(x,\, z_{0}-\delta)\\
(\frac{\partial_{z}F(x,z)}{\epsilon(x,z)})\mid_{z=z_0+\delta}& \equiv &(\frac{\partial_{z}F(x,z))} {\epsilon(x,z)})\mid_{z=z_0-\delta}
\end{eqnarray}
identically for $0\le x\le\Lambda$ and for $\delta\rightarrow0$,
for H-polarisation and likewise for E-polarisation dropping the term
$1/\epsilon(x,z)$. Inserting the expansion (\ref{eq:planewave})
leads to linear equations between the expansion coefficients $a_{n}^{(j)\pm}$
of the upper layer $j$ and $a_{n}^{(j-1)\pm}$ of the layer $j-1$
underneath. The fact that eigenmodes in the upper (lower) layer have
discontinuities at positions $x_{j}$ where the eigenmodes in the
lower (upper) layer are analytic, leads to a Gibbs phenomenon at $x=x_{j}$,
when eigenfunctions in the lower layer are expanded in terms of eigenfunctions
in the upper layer. This gives rise to slow convergence and thus a
large number of modes have to be included if the optical behaviour
at such structures needs to be known accurately. This is the case
even if the eigenmodes are known analytically. For H polarisation
the convergence is substantially slower than for E polarisation because
both the second and the first derivative in the direction perpendicular
to the interface are discontinuous, whereas for E polarisation only
the second derivative is discontinuous.

The substantial reduction in modes by fully exploiting
symmetry and the stable implementation of the boundary value problem
following \cite{morf1995exponentially} allows us to achieve the high
resolution necessary to obtain reliable absorption spectra for such
structures.

\paragraph*{Anti-Reflective Coating}

All the structures in this paper are calculated with an idealised
artificial antireflective coating that almost completely eliminates
reflection of the incident light at the top interface of the structure
structure. We choose this setting, because we wish
to describe an optimum limit for the absorption $A_{int}$ within the semiconductor
layer, while the light is incident from outside. The antireflective
coating used in our calculations consists of a homogeneous layer
with an artificial material AR{*} whose refractive index is chosen
wavelength dependent $n^{*}(\lambda)$=$\sqrt{\Re n_{a-Si}(\lambda)}$,
where $n_{a-Si}$ is the refractive index of a-Si in
the layer below, and the $\Re$ symbol stands for the real part. We
take the antireflection coating to have a thickness corresponding
to $\frac{\lambda}{4n^{*}(\lambda)}$. Of course this is very
artificial. However there exist realizations of antireflective
structures which lead to similarly low reflexion
losses \cite{heine1995submicrometer},\cite{heine1996coatedsubmicron}.
The broadband antireflection structure described therein consists
of a rectangular grating with an additional coating with a lower refractive
index than the grating below.

\section{Results}

\subsection{rectangular gratings\label{sub:rectangular-gratings}}

Here, we look at gratings consisting of only one grating layer. Furthermore,
we use an idealised, absorption free silver reflector whose permittivity
$\epsilon(\lambda)$ is given by the real part of the permittivity
of bulk silver, cf. Appendix. 

\begin{figure}[ht]
  \centering
  \includegraphics[width=7cm]{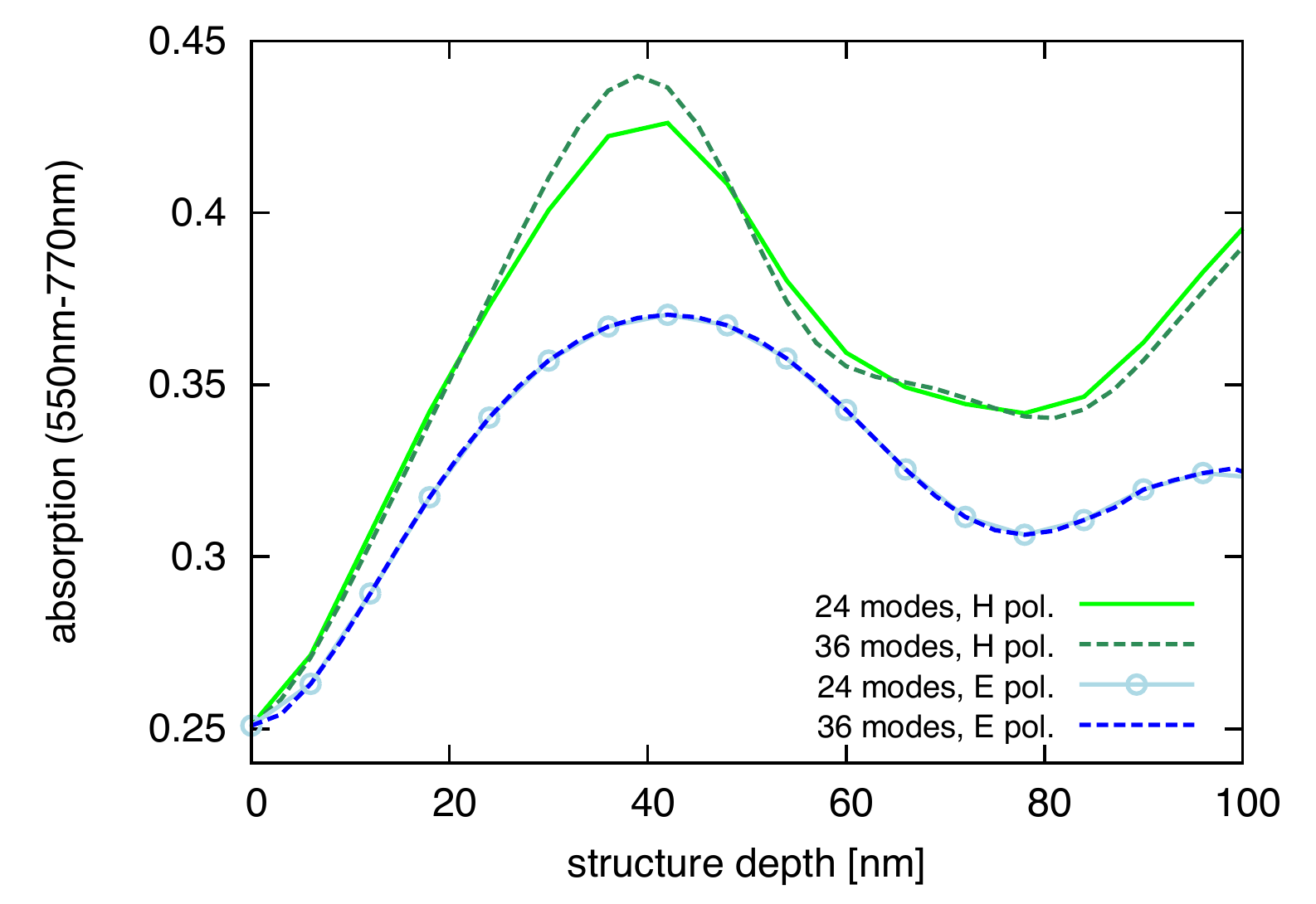}
  \vskip -2mm
  \caption{Depth dependence of absorption $A_{int}$ for rectangular grating with
    period $\Lambda=441$ nm for maximum absorption of unpolarised
    light and an a-Si layer with thickness $h_{mean}=200$ nm.} 
  \label{depthdependence-rect}
  \vskip -3mm
\end{figure}

The rectangular gratings are constructed like the ones shown in Figure
\ref{sym-antisym}, with a width of the metallic domain $b_{3}=0.5\Lambda$, and
$h_{2}=0$ and $h_{mean}=200$ nm, which stays constant while varying the
grating layer thickness. Thus, the amount of semiconductor material
is kept fixed. We also limit ourselves to grating depths that are at most
about one third of the semiconductor layer thickness. Figure \ref{depthdependence-rect}
shows the depth dependence of the absorption $A_{int}$ in a rectangular structure
at the optimum period for the first absorption maximum of unpolarised
light. In Figure \ref{depthdependence-rect} and the following similar
figures, an integrated absorption of $100$\% corresponds to the absorption of all
photons in the spectral range of $350$ nm up to $770$ nm, weighted
with an AM-1.5 spectrum. We limit the spectral range of our studies
to $550-770$ nm, where light trapping is most helpful, and the contribution
of this region to the total absorption is limited to $62.6$\% for
infinite thickness. 
Figure \ref{depthdependence-rect} shows that the maximum absorption $A_{int}$
for H parallel polarisation in the limited range of $550-770$ nm is almost $44$\%, whereas for E parallel
polarisation it is $\approx 37$\%, and the optimum depth for both polarisations
is close to $40$ nm. The figure also shows that the E parallel polarisation
is fully converged, whereas for H polarisation the value is between
$44.0$\% and $43.2$\%. To these values one may add a contribution
of 38\% for full absorption in the wavelength range from 350-550nm,
if reflection is suppressed. Thus, the total absorption $A_{int}$
amounts to 62\% for an ideal planar reflector, and if an ideal
reflector in the shape of this rectangular grating is used, it is 
75\% for E polarisation and between 81.2 and 82.0 percent for H polarisation.
\begin{figure}[h!]
  \centering
  	\includegraphics[width=6.6cm]{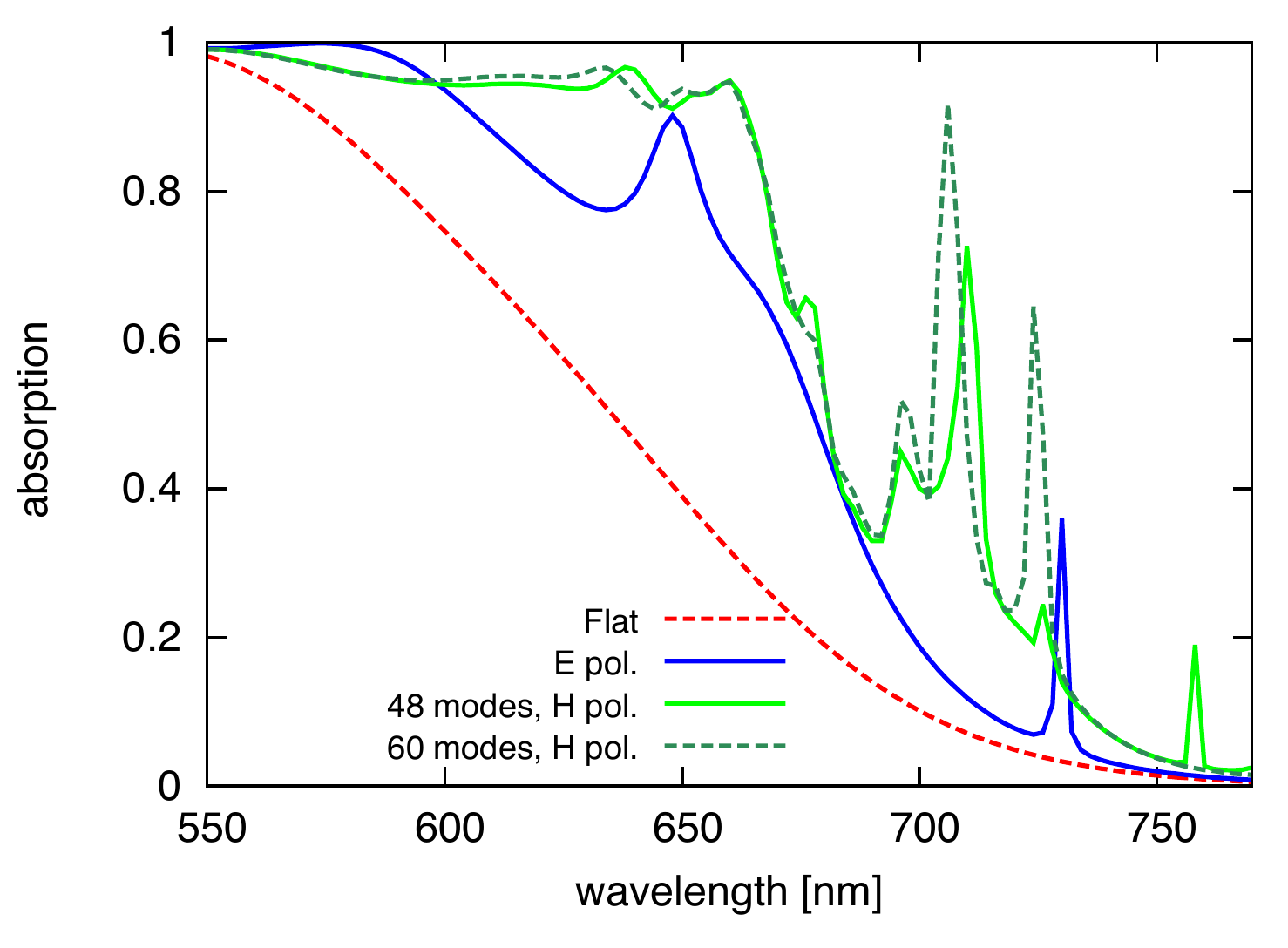}
  	\vskip -2mm
  	\caption{The absorption spectrum of the rectangular grating with $h_{mean}=200$
  	  nm, a period $\Lambda=401$ nm, and $36$ nm grating depth.}
  	\label{spectrum-rect}
  	\vskip -2mm
\end{figure}
\begin{figure}[ht]
  	\includegraphics[clip,width=7cm]{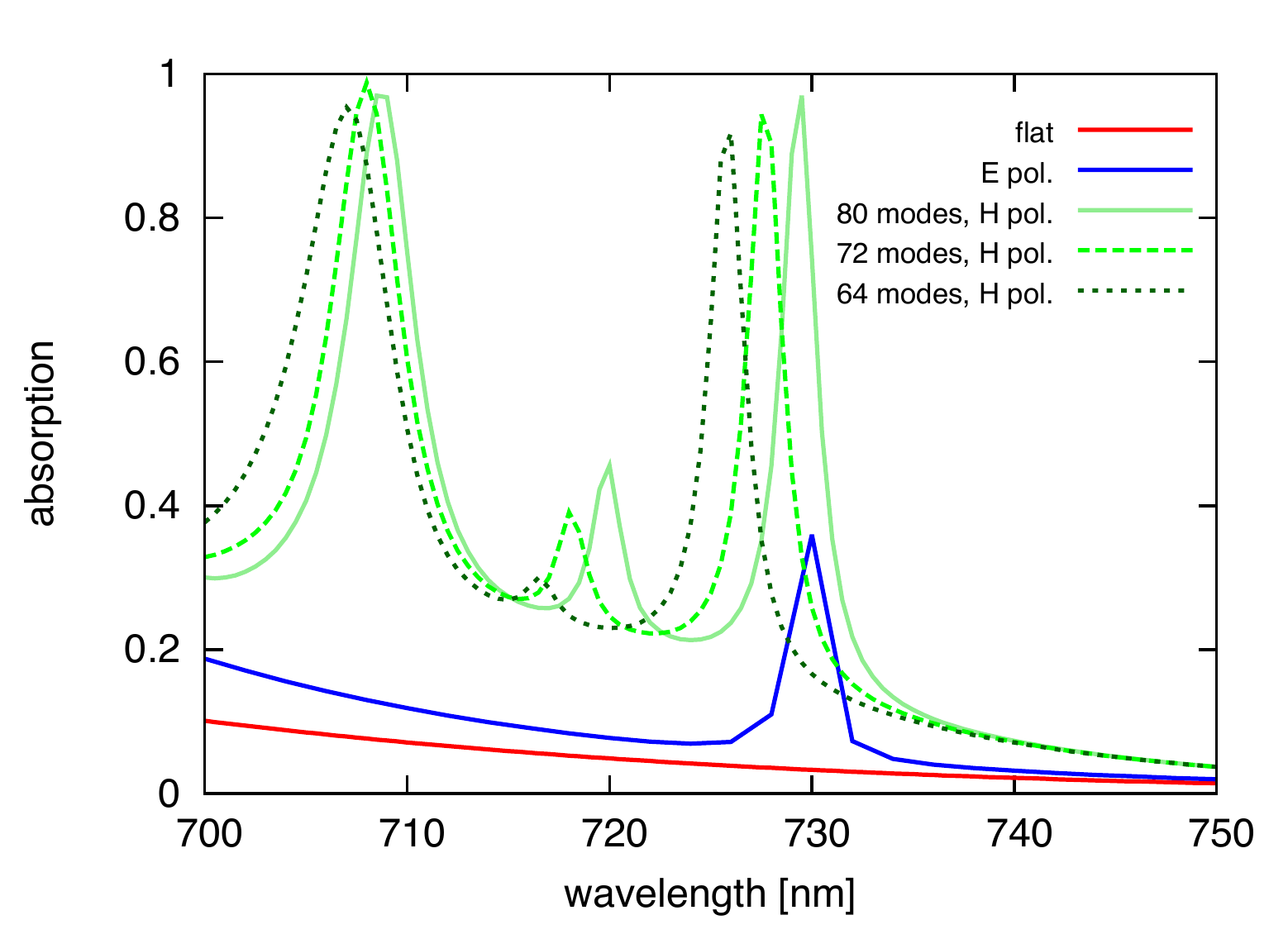}
  	\vskip -3mm
  	\caption{Absorption peaks in E-polarisation converged,
    	in H-polarisation they shift with N, keeping the integrated
        absorption fixed within 0.2\%.}
  	\label{spectrum-zoom}
  	\vskip -4mm
\end{figure}

The spectrum in Figure \ref{spectrum-rect} of the absorption shows
that the grating leads to a spectral shift of the absorption edge,
and the H parallel polarisation shows significantly more absorption,
especially in the infrared region. We observe that even for weakly
absorbing semiconductor layers, the strong resonances can lead to
a noticeable increase in absorption. Note that Figure \ref{spectrum-rect}
shows that the region between $700$ and $750$ nm is highly sensitive.
This behaviour is further investigated in Figure \ref{spectrum-zoom},
which shows spectral mode dependence of high order calculations. Note
how mostly the resonances move to higher wavelengths when increasing
the number of modes for the calculation, such as the one at $708$
nm. The integrated absorption values for H-polarisation vary little
for different mode numbers, even for small values of $N$, as can be
seen in Figure \ref{depthdependence-rect}, as has been mentioned above.
Due to the slow convergence we are unable to present absorption
spectra for H polarisation in the limit of large $N$.
On the contrary, for E polarisation our spectra are well converged.

Figure \ref{spectrum-zoom} also shows how the background of the absorption
spectrum, i.e. the broad absorption below the peaks clearly increases:
It may be taken as $\approx 0.3$ near the $708$ nm resonance, and
still around $0.2$ for the $728$ nm resonance. Such values would
demand much greater thicknesses $d$ of a-Si, namely
$d\approx700$ nm for the absorption value 0.3 at $\lambda=700$ nm
and even $d\approx1200$ nm for the value 0.2 at $\lambda=730$ nm,
if instead of the grating a planar reflector is used. Thus, the effective
thickness of the a-Si layer is increased by a factor
of 3.5 at $\lambda=700$ nm and about 6 at $\lambda=1200$ nm.
\begin{figure}[ht]
  \centering
  \includegraphics[width=6.6cm]{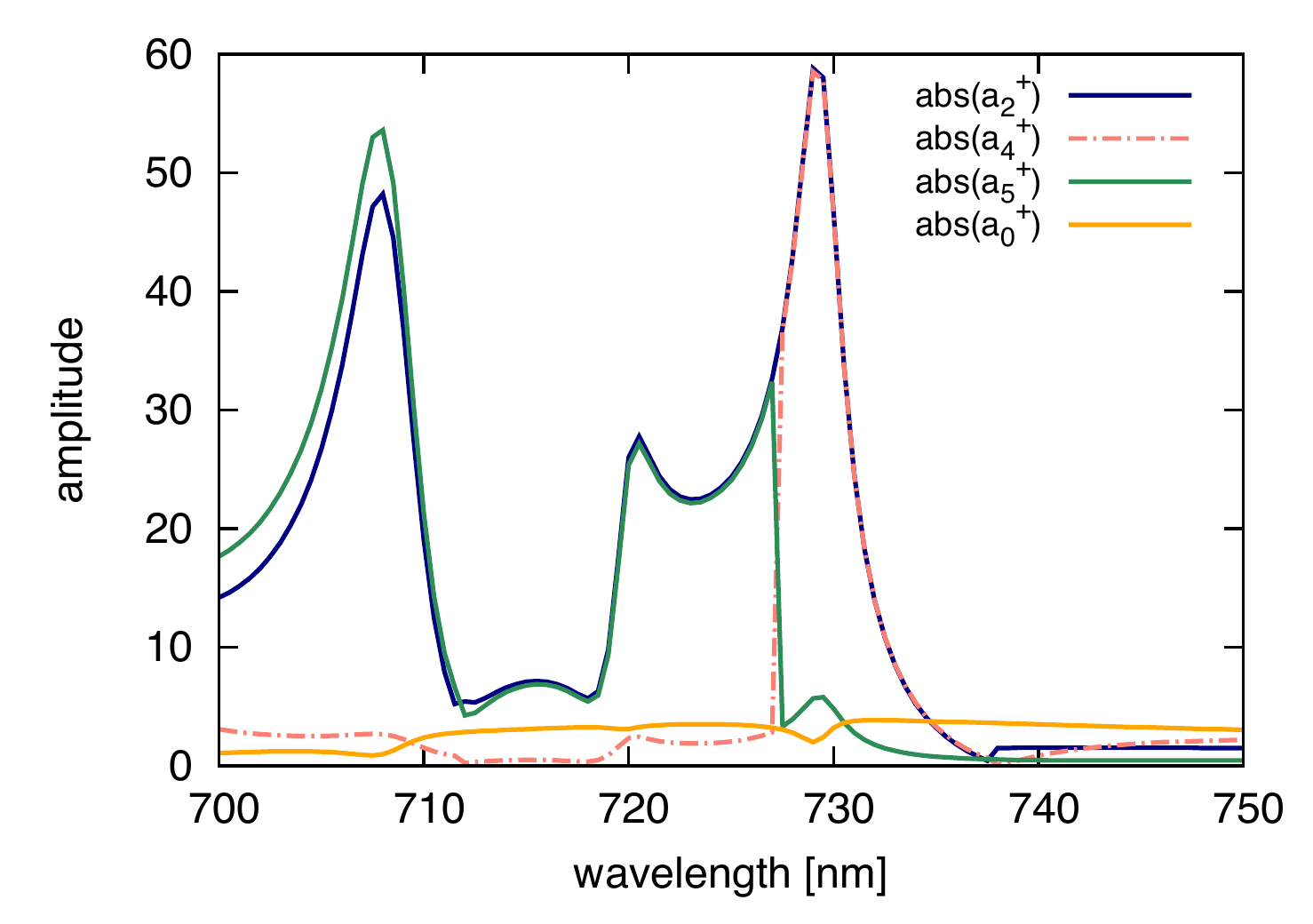}
  \vskip -3mm
  \caption{Spectrum of strongly excited modes $2$, $4$ and $5$ and of mode $1$ for comparison.}
  \label{spectrum-eigenmodes}
  \vskip -3mm
\end{figure}

In the following, we take a closer look at the eigenmodes of the rectangular
grating for which we have shown the spectrum in Figure \ref{spectrum-zoom}.
Figure \ref{spectrum-eigenmodes} shows the amplitudes
$a_{k}$ for the eigenmodes of the Helmholtz equation, sorted according
to increasing imaginary part of the eigenvalue implying shorter decay
length in the z-direction. The amplitudes are
shown at the interface between the grating layer and the metallic
reflector, and they result from solving the multilayer boundary value
problem, with an incident amplitude $a_{0}^{-}=1$, as defined in equation (\ref{F1}). Interestingly,
the second and fourth eigenmodes are almost 60 times as much excited as
the incident light, corresponding to a very high photon density near
the bottom interface. We expect this excitation to become much weaker
if losses in the metallic reflector are included. The fourth
and fifth eigenmodes switch order near  $\lambda=727$ nm.
\begin{figure}
  \centering 
  \includegraphics[width=7cm]{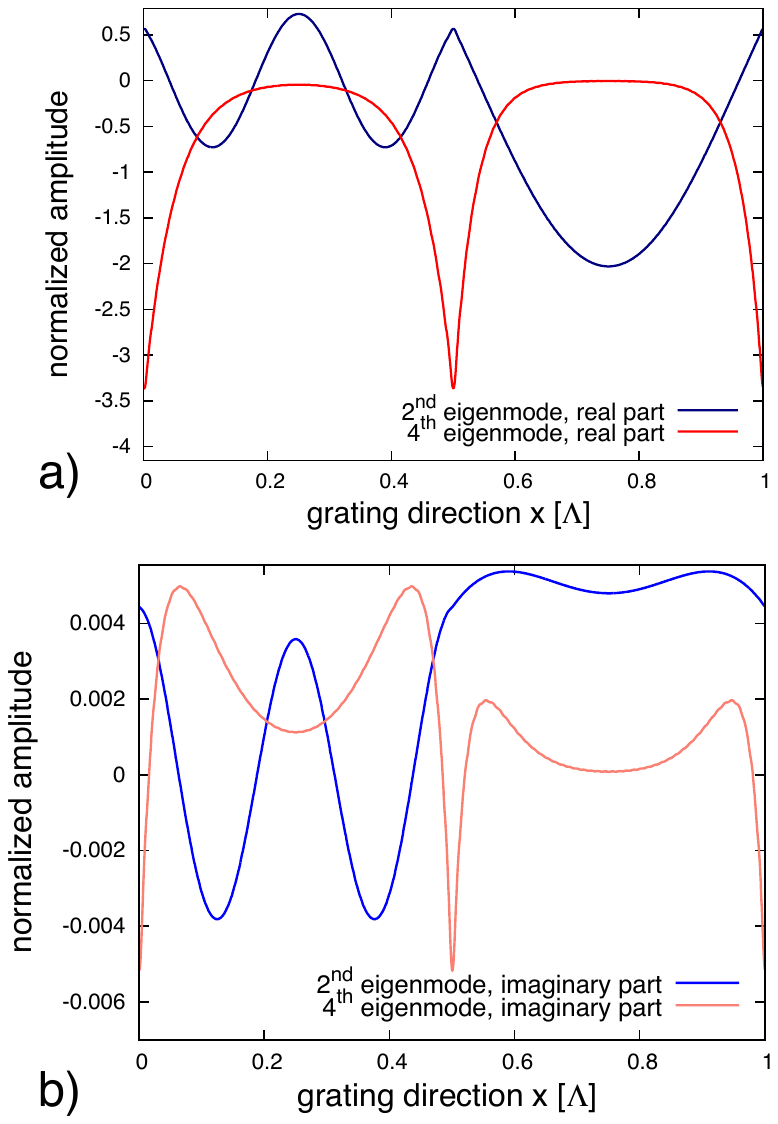}
  \caption{The strongly excited $2^{nd}$ and $4^{th}$
    eigen-functions at $\lambda=729$ nm. a) real, b) imaginary part.}
  \label{fig:spatial-eigenmodes}
  \vskip -3mm
\end{figure}

In Figures \ref{fig:spatial-eigenmodes}, we show the spatial behaviour
of the strongly excited eigenmodes. Note how both modes display a
sharp discontinuity at the interface between the two materials, as
required by the boundary conditions. The structure of these modes
shows clearly why the discontinuity of the first derivative gives
rise to a strong Gibbs phenomenon leading to the excitation of evanescent
waves of high order and consequently slow convergence as a function
of the number of modes included in the expansion of the electromagnetic
field. The comparatively small imaginary part in these eigenfunctions
is the result of the rather weak absorptivity of a-Si
at a wavelength of 729nm. We also note that a Fourier expansion of
these eigenfunctions will not allow to correctly compute the absorption
due to the 4-th eigenmode that mainly occurs in the vicinity of the
interface. Here, the derivative of the imaginary part shows a dramatic
discontinuity. Thus a Fourier expansion will converge only slowly in
its vicinity. In
view of the very large amplitude of this mode, the exact treatment of
the boundary conditions at the a-Si-AG$^*$ interface as implented in
our work is of particular relevance.

\subsection{Mimicking sinusoidal gratings\label{sub:Mimicking-sinusoidal-gratings}}

\begin{figure}%[hb]
  \centering
  \includegraphics[width=6.5cm]{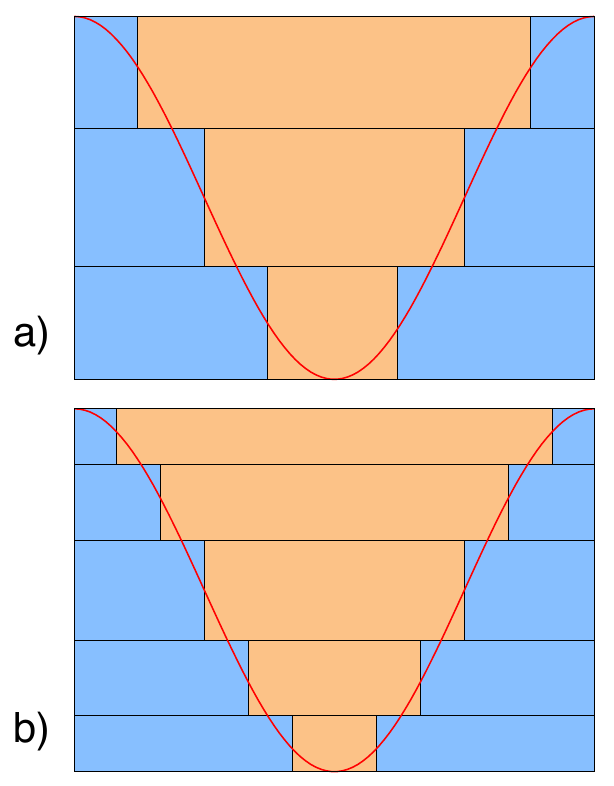}
  \vskip -2mm 
  \caption{Illustration of 3 (a) and 5 (b) layer staircase gratings mimicking a sinusoidal grating.}
  \label{pseudo-sinus-sketch}
\end{figure}

Here, we want to study more complex structures, in particular we are
interested in structures that mimic a sinusoidal grating. For that,
we want an optimised grating structure, which is not defined by a
number of parameters, but uniquely defined by a single parameter.
This should facilitate changing the structure depth without altering
the grating characteristics. To that end, we construct our grating
structure as follows: We fix the layer thickness $h_{j}$ and interface
positions $x_j$ by requiring that in each layer the excess area of the
lamellar approximation above the sine cancels the missing area
underneath it, and that the sum of excess areas is minimal and equal
to the sum of missing areas. This minimization leads to a unique
solution for $x_j$ and $h_j$ depending only on the number of layers
$n$, which we will refer to as pseudo-sine. The amount of
semiconductor material is then identical
for sine and pseudo-sine, see Figure \ref{pseudo-sinus-sketch}. 

%Here, we illustrate the cases of 3 and 5 layer grating versions
%mimicking sinusoidal gratings, with the property that the amount of
%semiconductor material is independent of grating depth and the areas
%of positive and negative deviations are all minimised and equal.

\begin{figure}[ht]
  \centering
  \includegraphics[width=6.3cm]{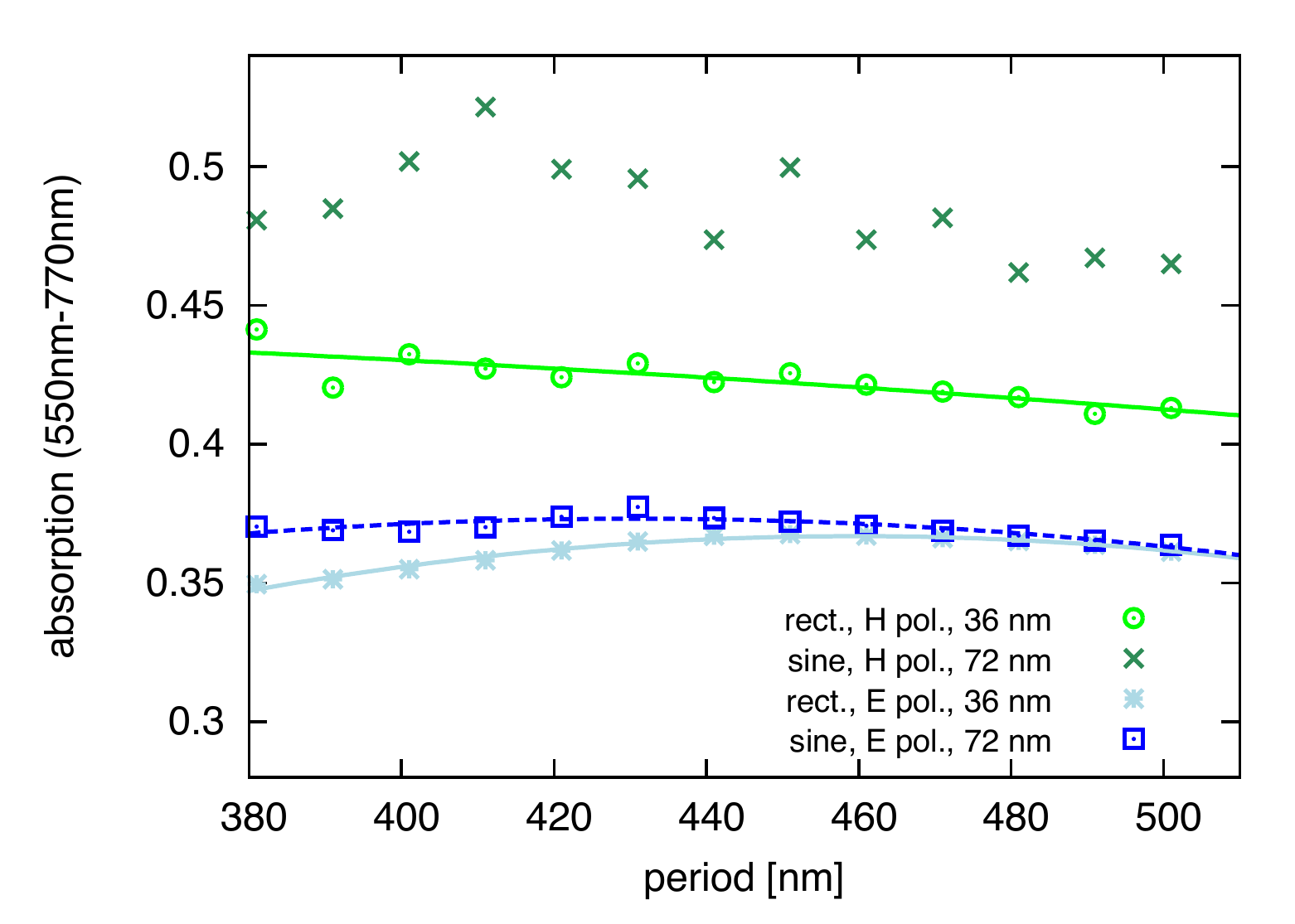}
  \caption{Grating period dependence of integrated absorption:
    rectangular grating with depth $h=36$ nm  vs. a $5$-step
    sine grating with $h_{tot}=\sum_j{ h_j}=72$ nm total depth.
    At these depth values, integrated absorption is close to maximum. 
    a-Si layer thickness $h_{mean}=200$ nm.}
  \label{perioddependence-rect}
\end{figure}

\begin{figure}[ht]
  \centering
  \includegraphics[width=6.8cm]{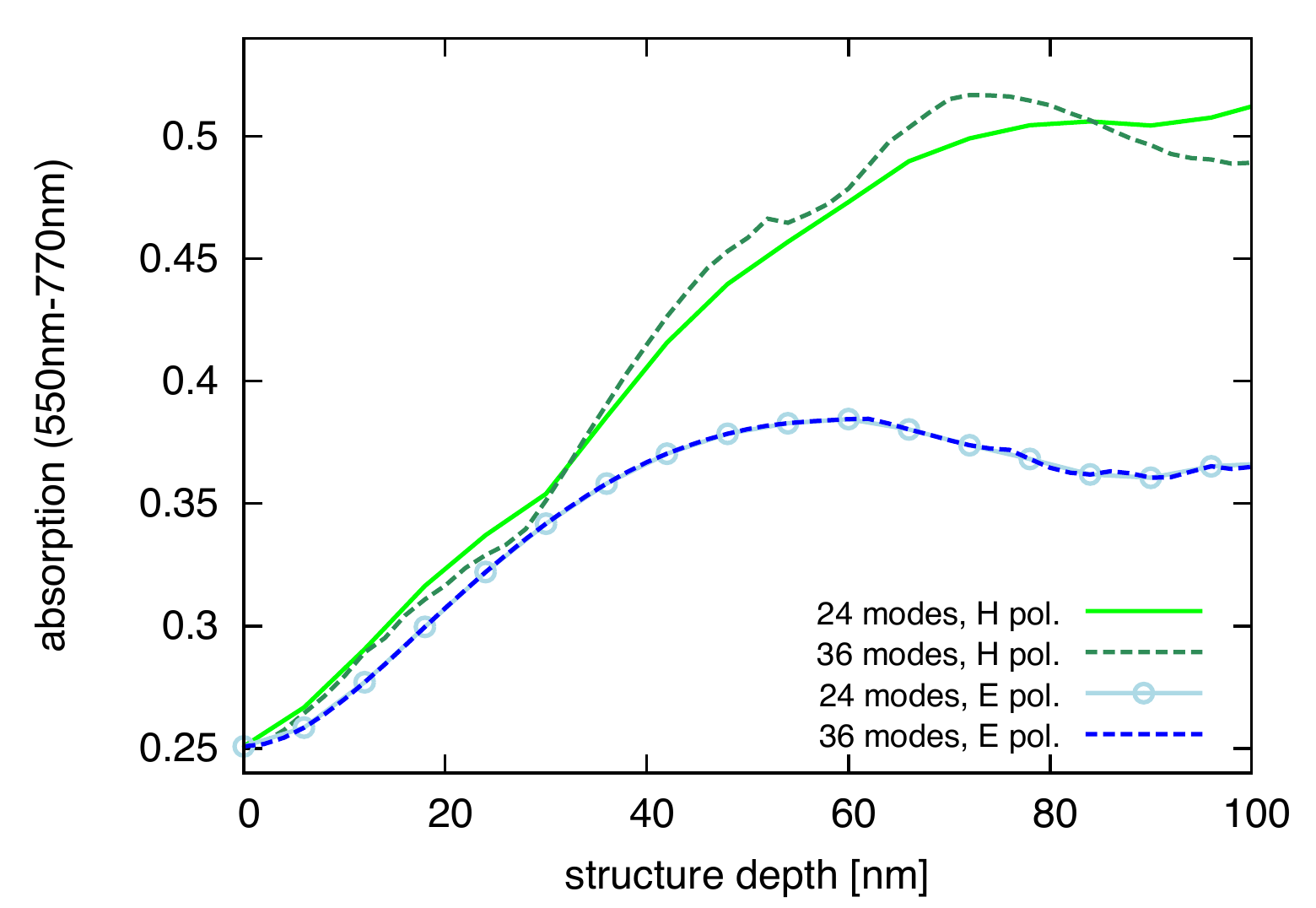}
  \caption{Depth dependence for a 5 step pseudo-sine grating with period $\Lambda=421$
    nm.}
  \label{depthdep-sine}
\end{figure}

Figure \ref{perioddependence-rect} shows the integrated absorption $A_{int}$
for the interval between $550$ nm and $770$ nm, and how the first
absorption maximum for both rectangular and pseudo-sine gratings only
depends weakly on the period $\Lambda$. In particular, we can see
that the 5-step pseudo-sine offers more absorption, and that in both
cases, the H polarisation is more readily absorbed. In Figure \ref{perioddependence-rect},
the depth of $36$ and $72$ nm for the rectangular and for the pseudo-sine
structure respectively are chosen close to the first absorption maximum
for each structure type at around $40$ and $70$ nm as can be seen
in Figures \ref{depthdependence-rect} and \ref{depthdep-sine}.

Figure \ref{depthdep-sine} shows that the integrated absorption
depends weakly on the number of modes used for the calculation, and
in particular that for the E parallel polarisation, $24$ modes are
sufficient to achieve convergence, whereas similar curves for
H parallel polarisation are already in good agreement, but still varying
for deep structures. It also shows a significant difference in optimum
depth between polarisations, unlike for rectangular gratings, and
that H parallel polarised light can be absorbed more effectively.
When further comparing the depth dependence in Figure \ref{depthdep-sine}
with Figure \ref{depthdependence-rect}, note that the optimum depth
for a pseudo-sine structure is almost twice the optimum depth of the
corresponding rectangular structure, and the optimum for the pseudo
sine structure is much flatter with respect to varying depths. 

\begin{figure}%[ht]
  \centering
  \includegraphics[width=6.8cm]{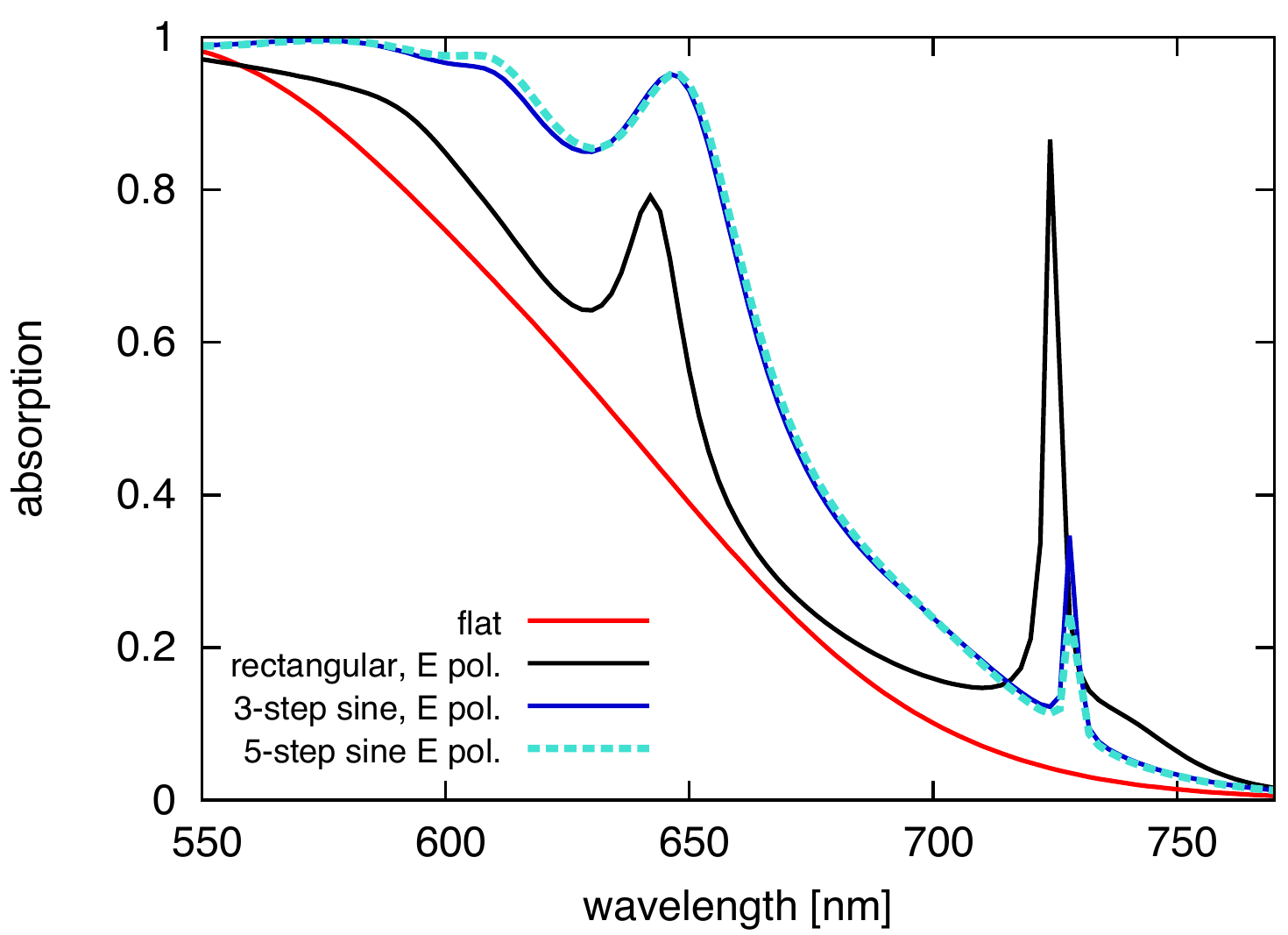} 
  \caption{Spectrum of 3 and 5-step pseudo-sine vs rectangular and flat reference, $N=36$ modes for E parallel incidence}
  \label{spectrum-sine-E}
\end{figure}

\begin{figure}%[ht]
  \centering           
  \includegraphics[width=6.8cm]{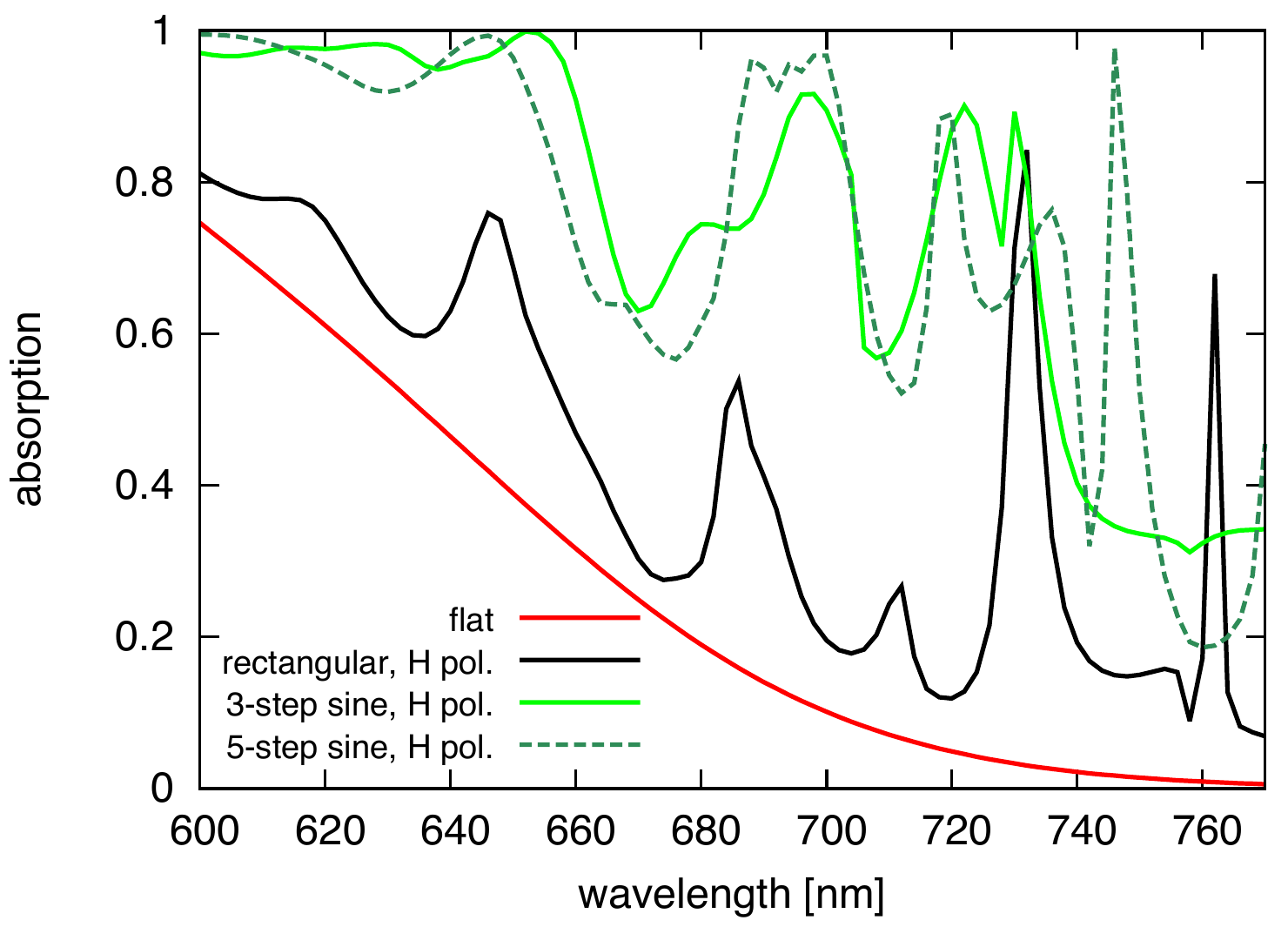}
  \caption{Spectra for H parallel incident light, same gratings as in Figure \ref{spectrum-sine-E}}
  \label{spectrum-sine-H}
  \vskip -4mm
\end{figure}

The spectra in Figure \ref{spectrum-sine-E} and \ref{spectrum-sine-H}
show that the pseudo-sine gratings offer more absorption over the
integrated spectral range when compared to the rectangular gratings,
both by increasing the background absorption in the red and near
infrared, and by exciting additional resonances. 
We note that the background absorption of the pseudo-sine grating
without taking any peaks into account takes values of about 0.6
at $\lambda=680$ nm and about 0.4 at $\lambda=730$ nm. A device
with a planar reflector would have to have an a-Si thickness
of about $1000$ nm at $680$ nm wavelength, and even $2700$ nm at
$730$ nm wavelength, i.e. 5 and 13 times the actual thickness of the
device with the pseudo-sine grating.

For the pseudo-sine, the optimum parameters in terms of depth and
period for the two polarisations are further apart when compared to
a rectangular grating. The spectral comparison of 3-step and 5-step
pseudo-sine gratings with period $401$ nm and depth $72$ nm in Figure
\ref{spectrum-sine-E} reveals that the spectra are almost identical
for the E parallel polarisation. Thus the absorption spectra are almost
independent of the number of layers chosen for the pseudo-sine approximation.
This shows the validity of our geometrical argument for the unique
choice of the pseudo-sine approximation. On the other hand, the same
spectral comparison for the H polarisation in Figure \ref{spectrum-sine-H}
shows differing spectra for the 3 and 5-step pseudo-sine gratings,
but they do share common spectral features, explaining the good convergence
of the integrated absorption. Because both the mode number and the
number of layers included are very limited, the pseudo-sine spectrum
is not converged and must not be confused with the spectrum for a
true sine grating, for which other methods such as the one pioneered
by Chandezon et al. \cite{chandezon1980original} will be required.
The 5-step pseudo-sine grating shows a maximum integrated absorption
in H polarisation of about 52\%, while for unpolarised light it is
given by the mean value from E- and H-polarisation and is about 45\%. 
These values include the spectral range from 550-770nm. Adding
the contribution of 38\% from the wavelength range 350-550 nm, where
light-trapping is not needed, we obtain a total of 90\% for H
polarisation and 83\% for unpolarised light.

\subsection{asymmetric gratings\label{sub:asymmetric-gratings}}

In this section we show the benefits of asymmetry in
the grating structure, as shown in Figure \ref{sym-antisym}. All these
results refer to the case of E-polarisation. The calculations for
H-polarisation have not yet achieved the reliability that we require in
this work. Further developments will be necessary to overcome these
difficulties.

\begin{figure}[ht]
  \centering
  \includegraphics[width=6.8cm]{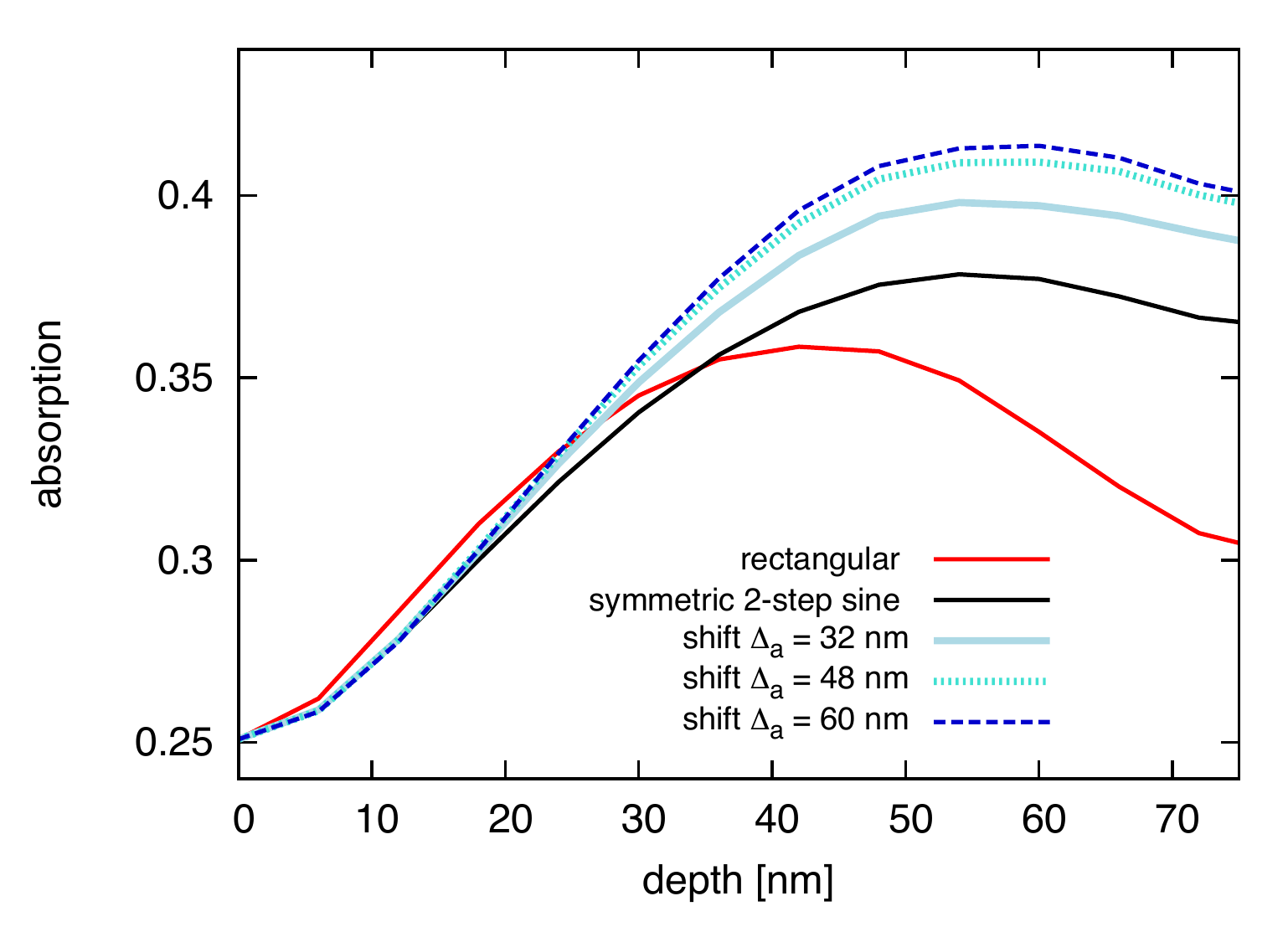}
  \vskip -2mm
  \caption{Depth dependence of absorption $A_{int}$ in E-polarisation, asymmetric grating, $39$ modes.}
  \label{Depth-asy}
\end{figure}

In Figure \ref{Depth-asy}
we have calculated the depth dependence of a two-step pseudo-sine
grating with period $\Lambda=401$ nm where one layer is shifted by
$\triangle_{a}$ from the symmetry axis, using 39 modes which is sufficient
to display the high resolution limit behaviour. Figure \ref{Depth-asy}
shows that the asymmetry improves the absorption from around $37$\% maximum to around $41$\%,
and that the benefit of the asymmetric shift saturates if $\Delta_a$ is increased beyond $0.1\Lambda$.
% of the grating period $\Lambda$.

\begin{figure}%[ht]
  \centering
  \includegraphics[width=6.8cm]{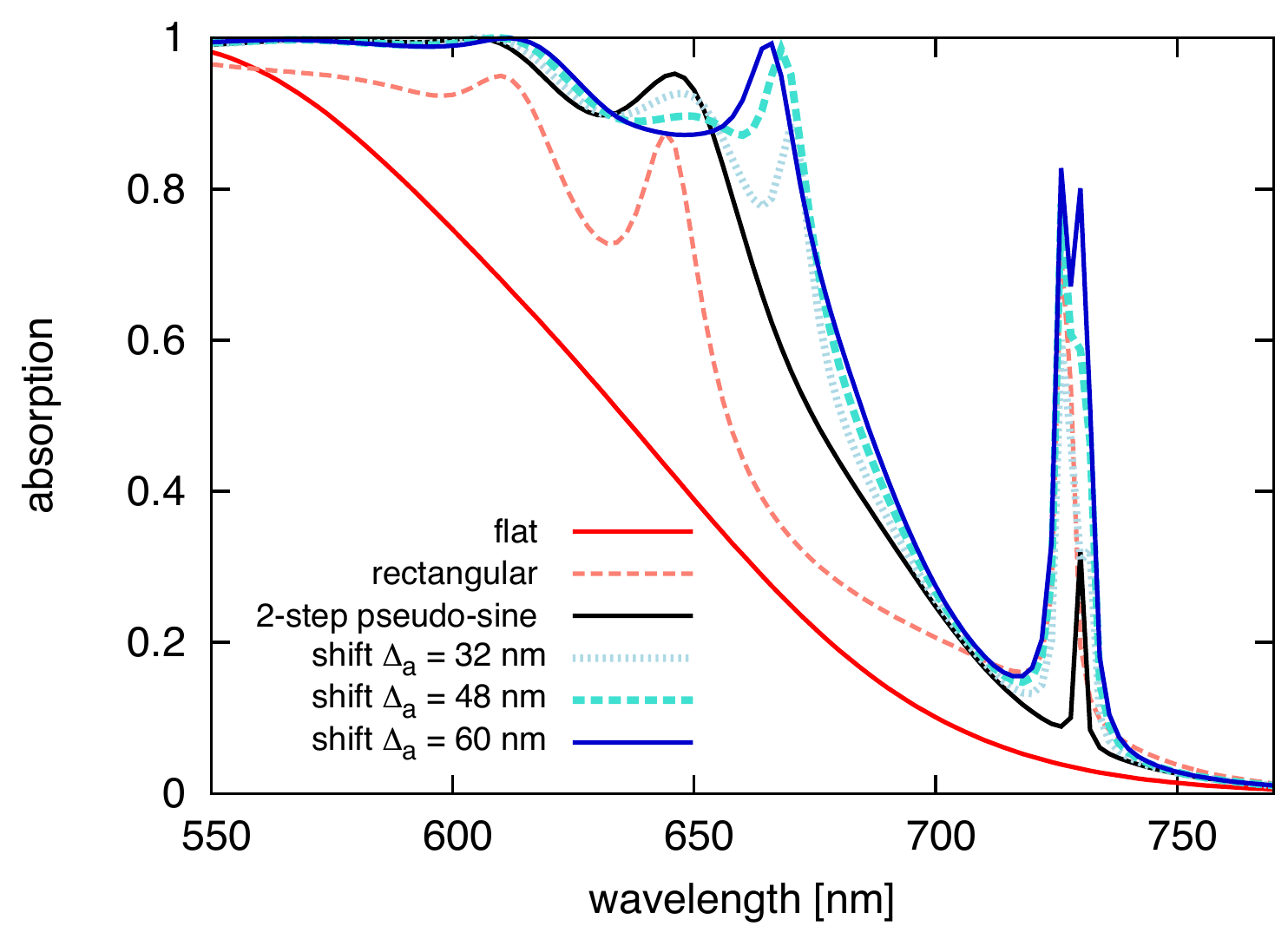}
  \vskip -2mm
  \caption{Spectra of asymmetric and rectangular gratings for a grating depth of 60 nm.}
  \label{Spectrum-Asy-a}
\end{figure}
\begin{figure}[ht]          
  \centering
  \includegraphics[width=6.8cm]{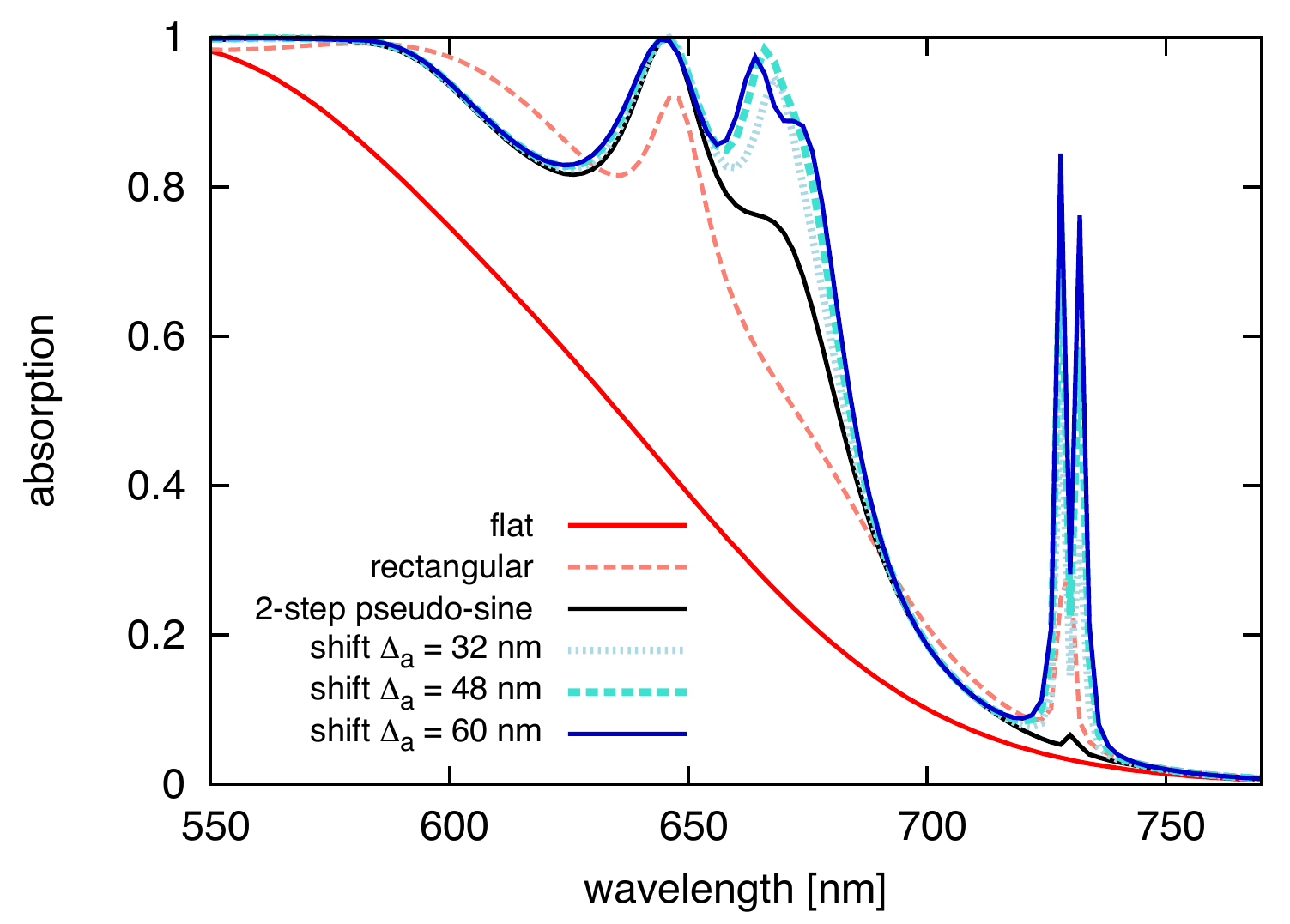}
  \vskip -2mm 
  \caption{Spectra asymmetric and rectangular gratings for a grating depth of 42 nm.}
  \label{Spectrum-Asy-b}
  \label{Spectrum-Asy}
\end{figure}
\begin{figure}%[hb!]
  \centering
  \includegraphics[width=6.8cm]{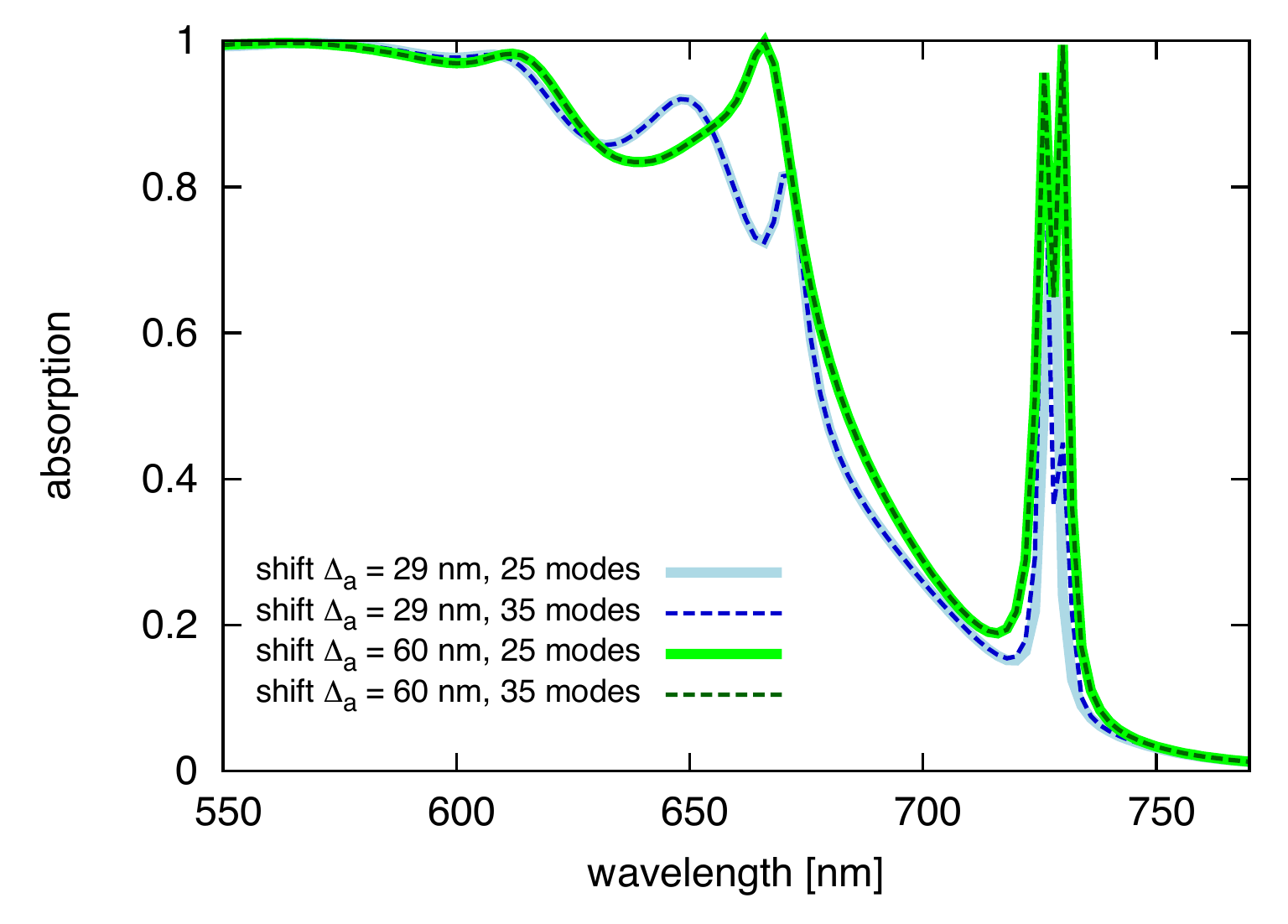}
  \vskip -2mm
  \caption{A comparison of the spectra of asymmetric structures
    for E parallel polarisation}
  \label{asy-modes}
\end{figure}
Figure \ref{Spectrum-Asy-a} shows that for a structure with depth
$60$ nm, the asymmetry not increases the background absorption
into the red and near infrared, but also introduces additional resonances
around $670$ nm and around $730$ nm, which correspond to the excitation
of two antisymmetric eigenmodes of the grating. For comparison, the
spectrum at a depth of $42$ nm in Figure \ref{Spectrum-Asy-b} shows
absorption resonances at the same wavelengths. To the maximum absorption
of $41$\% observed for E-polarisation with this  asymmetric grating
integrated over $550\le\lambda\le 770$ nm we can add the absorption
of 38\% from $350\le\lambda\le 550$ nm, where light-trapping is not needed,
and obtain a total absorption of 79\%, quoted in the Abstract. Finally,
Figure \ref{asy-modes} shows the absorption in E-polarisation of a grating
with period $401$nm and depth $66$ nm calculated with $N=25$ and 35 modes. Despite
large asymmetry, the spectra have well converged.

\section{Discussion}
We have investigated how much light can be trapped in a layer of a-Si
with thickness 200 nm, which is typically used for a-Si solar cells,  on top 
a metallic surface relief grating with one-dimensional (1-d) lamellar
geometry. In order to obtain upper
limits for the absorption, we have made the idealizations of a
loss-free metallic reflector and a perfect antireflection coating.
We have used a modal method with exact eigenfunctions
to calculate accurate absorption spectra of such gratings.
The absorption spectra for E polarisation of a sine-shaped diffraction grating
approximated by lamellar gratings with 3 and 5 layers turn out almost identical,
which is an interesting result. For such sine-like gratings, we have found
that the limit of absorption is $A_{int}$ is about 90\% for H-polarisation and
about 75\% for E-polarisation, compared to $62$\%
for a planar reflector. We have also shown that an
asymmetry in the grating structure generates additional peaks in
the absorption spectrum and can thus increase the absorption.

In Section 3.1, we have shown that {\bf exact} calculation of eigenmodes
is particularly important as significant absorption occurs near
interfaces where derivatives of the fields are
discontinuous, see Figure  \ref{fig:spatial-eigenmodes}.

While we have shown that in all cases studied, light-trapping in
H-polarisation is significantly more effective than in E-polarisation,
calculating fully converged spectra for H-polarisation remains
difficult. The results presented in this paper have been made possible
by making full use of the symmetry properties of the structures studied.
One may hope that two-dimensional surface
relief gratings of similar lamellar geometry will allow unpolarised
light to be absorbed as effectively as H-polarised light is absorbed
with the 1-d gratings, reported here.
As already pointed out, our
results are idealised in the sense that absorption
in the metallic layer is suppressed. We expect that the peak
height of absorption resonances will be
strongly reduced by experimentally available metallic reflectors.
On the other hand, one may hope that the background absorption below the peaks
that is observed for all studied gratings, will still contribute significantly.

\section*{Acknowledgments}

We acknowledge the financial support of this work by the Swiss Federal
Office of Energy, and by the Paul Scherrer Institute. We also
like to thank F.J. Haug and H.P.Herzig for helpful discussions and
advice.
\section*{References}
\bibliographystyle{iopart-num}
\bibliography{referenzen1}
\subsection*{Appendix: permittivities of amorphous silicon and idealised silver}

For our calculations we have employed dispersion data for amorphous
silicon supplied by the PV-LAB of EPFL, Lausanne, Switzerland. To facilitate
test calculations by the interested reader, we have used fit formulae
to parameterize those measured data, valid for wavelengths $0.4 \le\lambda\le 0.77$ $\mu$m.

Denoting the complex refractive index by $\hat{n}=n+i\, k$, and
measuring wavelength in $\mu$m, our fit functions for a-Si are given by
\begin{eqnarray}
n(\lambda) & = & 0.466+\lambda(29.923+\lambda(-59.388+34.092 \lambda))\\
%log_{10}k(\lambda) & = & 6.695+\lambda(-35.226+\lambda(68.444-50.832 \lambda))
k(\lambda) & = &10^{ 6.695+\lambda(-35.226+\lambda(68.444-50.832 \lambda))}
\end{eqnarray}
The particular form for the imaginary part $k(\lambda)$ was chosen
in order to properly describe the behaviour of the absorption constant
in the infrared part of the spectrum.

Likewise, we use for our idealized $Ag^{*}$ a real dielectric constant
$\epsilon(\lambda)$ parameterized by
\begin{equation}
\epsilon(\lambda)=5.53900-5.26855\lambda-45.38207\lambda^{2}
\end{equation}
where the wavelength $\lambda$ is again measured in micrometers.
Note that these fit functions are valid only for $400<\lambda<770$ nm.
We have used them for $550<\lambda<770$ nm. 

%\lipsum
\end{document}